\documentclass[prc,10pt,final,notitlepage,oneside,onecolumn,%
               superscriptaddress,nobibnotes,nofootinbib,showpacs]{revtex4}
\usepackage{graphicx}

\newcommand{\ds}{\displaystyle}
\newcommand{\mc}[2]{\multicolumn{#1}{|c}{#2}}

\begin{document}
\title{DECAY PROPERTIES OF THE DIPOLE ISOBARIC ANALOG RESONANCES}
\author{M.L.~Gorelik}
\author{I.V.~Safonov}
\affiliation{Moscow Engineering Physics Institute, 115409 Moscow,
Russia}
\author{M.H.~Urin}
\affiliation{Kernfysisch Versneller Instituut, University of
Groningen, 9747 AA Groningen, The Netherlands} \affiliation{Moscow
Engineering Physics Institute, 115409 Moscow, Russia}

\begin{abstract}
A continuum-RPA-based approach is applied to describe the decay
properties of isolated dipole isobaric analog resonances in nuclei
having not-too-large neutron excess. Calculated for a few
resonances in $^{90}$Zr the elastic E1-radiative width and partial
proton widths for decay into one-hole states of $^{89}$Y are
compared with available experimental data.
\end{abstract}
 \maketitle

\section{Introduction}
In nuclei having not-too-large neutron excess $N-Z=2T$ ($T$ is the
ground-state isospin value) the familiar giant dipole resonance
(GDR) is splitted into two components, characterized by the isospin
values $T_<=T$ and $T_>=T+1$. The GDR $T_>$-component (GDR$_>$) is
the isobaric analog of the charge-exchange (in the
$\beta^+$-channel) giant dipole resonance (i.e. of the GDR$^{(+)}$).
Thus, the GDR$_>$ presents the specific double giant resonance (see,
e.g. Ref.~\cite{Harakeh}). Because of Pauli blocking, the
GDR$^{(+)}$ is not strongly collectivized and contains along with
the main peak a few low-energy $1^-$-states of the
neutron-(proton-hole) type~\cite{Auerbach,Safonov}. These states is
a base for the GDR$_>$ low-energy part, which consists of a few
isolated dipole isobaric analog resonances (DIAR)~\cite{Safonov}.
One of these resonances in $^{90}$Zr ($E_x=16.28$ MeV) has been
identified in reactions with protons, electrons and photons (see
Ref.~\cite{Cosel} and references therein). The comprehensive study
of the mentioned DIAR with the $(e,e'p)$-reaction has been reported
in Ref.~\cite{Cosel}, where along with the form factor for two
values of the transferred momentum the elastic E1-radiative width
and partial proton widths for decays into one-hole states of
$^{89}$Y have been deduced for this resonance. Being motivated by
these experimental results, which are still not explained, in the
present work we attempt to describe the gross and (direct) decay
properties of isolated DIAR within a continuum-RPA(CRPA)-based
approach. The basic elements of such description are presented in
Sect.~\ref{crpa}. The applications to the DIAR in $^{90}$Zr are
given in Sect.~\ref{diar}, where the calculation results are
compared with available experimental data.

\section{\label{crpa}Continuum-RPA-based description of DIAR.}
\subsection{\label{diar_prop}DIAR properties}
The dipole isobaric analog resonances in a target nucleus $(Z,N)$
correspond to (2particle-2hole)-type excitations with a small
admixture of (proton-(proton-hole))-type states. Namely due to
this admixture, E1-decay of the DIAR into the target-nucleus
ground state and also (direct) proton decays into one-hole states
of the product nucleus $(Z-1,N)$ are possible. The description of
the DIAR properties is simplified, provided that the isobaric
analog state (IAS) is considered as the ``ideal'' one. The
``ideal'' IAS exhausts 100\% of the Fermi strength and its wave
function is generated by the Fermi operator ${\cal
T}^{(-)}=\sum_a\tau^{(-)}_a$. (The real IAR exhausts about 95\% of
the Fermi strength). In this approximation the wave function
$|s_>,1M\rangle$ of a GDR$_>$-component can be expressed via the
parent GDR$^{(+)}$-component wave function $|s^{(+)},1M\rangle$:
\begin{equation}
|s_>,1M\rangle=(2T_>)^{-1/2}{\cal T}^{(-)}|s^{(+)},1M\rangle.
\label{eq1}
\end{equation}
In the mean-field approximation, the ``ideal'' IAS excitation energy
equals the Coulomb displacement energy: $\Delta_C = (2T_>)^{-1}\int
U_C(r)n^{(-)}(r)d^3r$, where $U_C(r)$ is the mean Coulomb field,
$n^{(-)}(r)$ is the neutron-excess density of the parent nucleus
$(Z-1,N+1)$. (We assume that the IAS energy is independent of the
parent state).

Using Eq.~(\ref{eq1}), we express the GDR$_>$ strength function
(corresponding to the single-particle external field
$V^{(0)}(x)=(-1/2)\tau^{(3)}rY_{1M})$ via the GDR$^{(+)}$ strength
function (corresponding to the external field
$V^{(+)}(x)=\tau^{(+)}rY_{1M}$):
\begin{equation}
S^{(0)}_>(\omega)=(2T_>)^{-1}S^{(+)}(\omega^{(+)}_C), \label{eq2}
\end{equation}
where $\omega$ is the target-nucleus excitation energy,
$\omega^{(+)}_C=\omega-\Delta_C$~\cite{Safonov}. The parent-nucleus
dipole strength function $S^{(+)}(\omega^{(+)})$ can be calculated
within any CRPA-based approach (see e.g. Refs.~\cite{Auerbach,
Safonov}). The cross section of photo-absorption, accompanied by
GDR$_>$ excitation, is proportional to the strength function
$S^{(0)}_>(\omega)$: $\sigma^>_a(\omega)=B\omega S^{(0)}_>(\omega)$,
where $B=(16\pi^3/3)(e^2/\hbar c)$. This cross section, integrated
over an isolated DIAR (with the energy $\omega_s$),
$(\sigma^>_a){}^{int}_s$, determines the E1-radiative width for DIAR
decay into the target-nucleus ground state:
$\Gamma_{\gamma_0,s}=(1/3\pi^2)(\omega_s/\hbar c)^2
(\sigma^>_a){}^{int}_s$.

Being applied to an isolated DIAR, Eq.~(\ref{eq2}) allows one to
express the (one-dimensional) particle-hole DIAR transition density
via the corresponding parent-state transition density:
$\rho_{>,s}(r)=(2T_>)^{(-1/2)}\rho^{(+)}_s(r)$. In particular, this
transition density can be used for evaluation of  the DIAR form
factor, which determines the $(ee')$-reaction cross
section~\cite{Cosel}. Within the impoved plane wave approximation
the form factor equals:
\begin{equation}
|F_s(q)|^2 = 12\pi Z^{-2}{\cal Y}^2(Z)
\left|\int\rho_{>,s}(r)j_1(qr)dr\right|^2, \label{eq3}
\end{equation}
where ${\cal Y}(Z)=\ds\frac{2\pi
Z}{137}\left(1-\exp\left(-\frac{2\pi Z}{137}\right) \right)^{-1}$ is
the correcting factor (see, e.g., Ref.~\cite{Landau}), and $j_1(qr)$
is the dipole spherical Bessel function.

The basic quantity in CRPA-based description of DIAR proton decay is
the amplitude of the partial $(\gamma p)$-reaction, accompanied by
DIAR excitation. The poles in the energy dependence of this
amplitude correspond to DIAR, while the pole residue is proportional
to the product of the amplitudes of the elastic radiative and
partial proton widths. Thus, the joint analysis of the corresponding
calculated photo-absorption and $(\gamma p)$-reaction cross sections
allows one to evaluate the partial proton widths for a given DIAR.

\subsection{CRPA equations}
The CRPA expressions for the strength functions
$S^{(\pm)}(\omega^{(\pm)})$ and $(\gamma p)$-reaction amplitudes can
be obtained taking the CRPA equations in the form consistent with a
phenomenological version of the Migdal's finite Fermi-system theory.
As applied to description of charge-exchange giant resonances, these
equations are given in detail in Refs.~\cite{Moukhai,Gor1} and for
reader's convenience we use, in the main, the notations from these
references.

Of particular importance is the equation for the effective
single-particle external field
$\widetilde{V}^{(\pm)}(x,\omega^{(\pm)})$. The difference between
the effective field and the corresponding external field
$V^{(\pm)}(x)=V(r)Y_{LM}(\vec n)\tau^{(\pm)}$ is due to the
polarization effect from the particle-hole interaction taken for
description of the non-spin-flip excitations in the form:
$F_{p-h}\rightarrow 2F'(\tau^{(+)}\tau^{(-)'}+\tau^{(-)}\tau^{(+)'})
\delta(\vec r -\vec r')$. The effective-field radial part
$\widetilde{V}^{(\pm)}(r,\omega^{(\pm)})$ determines all the main
characteristics of a given charge-exchange giant resonance: the
strength function $S^{(\pm)}_V(\omega^{(\pm)})$, the
energy-dependent transition density
$r^{-2}\rho^{(\pm)}(r,\omega^{(\pm)})Y_{LM}$, and the nucleon-escape
amplitude $M^{(\pm)}_{V,c}(\omega^{(\pm)})$ ($c$ is the set of the
nucleon-decay-channel quantum numbers). In particular, the dipole
strength function $S^{(+)}(\omega^{(+)})$ of Eq.~(\ref{eq2}) can be
directly calculated via the radial part of the corresponding
effective field (see, e.g., Eq.~(1) of Ref.~\cite{Moukhai}). Here we
show the useful relationships, which have not been exploited in
Refs.~\cite{Auerbach,Safonov,Moukhai,Gor1}. The pole representations
of the effective-field radial part and strength function are valid
in the energy region of each collective neutron-(proton-hole)-type
state $|s^{(+)},1M\rangle$ with the excitation energy
$\omega^{(+)}_s$ below the neutron separation threshold:
\begin{equation}
\widetilde{V}^{(+)}(r,\omega^{(+)})\rightarrow \frac{2F'}{r^2}
\frac{d^{(+)}_s\rho^{(+)}_s(r)}{\omega^{(+)}-\omega^{(+)}_s+i0};
\qquad\qquad S^{(+)}(\omega^{(+)})\rightarrow
-\frac{1}{\pi}\mathrm{Im}
\frac{(d^{(+)}_s)^2}{\omega^{(+)}-\omega^{(+)}_s+i0}. \label{eq4}
\end{equation}
Here, $\rho^{(+)}_s(r)$ is the transition density of the
(stationary) collective state $|s^{(+)},1M\rangle$, and
$d^{(+)}_s=\int r\rho^{(+)}_s(r)dr$ is the dipole matrix element. In
particular, this matrix element determines the partial radiative
width $\Gamma_{\gamma_0,s}$ of the corresponding DIAR
(Subsect.~\ref{diar_prop}).

The transition-density elements, which define the transition density
$\rho^{(+)}_s(r)=\sum_{\nu\pi}\rho^{(+)s}_{\nu\pi}\chi_\pi(r)
\chi_\nu(r)$, can be found from the (formally, infinite range)
system of the homogeneous equations, which is usually used within a
discrete-RPA:
\begin{equation}
\rho^{(+)s}_{\nu\pi}=t^{(1)}_{(\nu)(\pi)}\frac{n_\nu-n_\pi}
{\varepsilon_\nu-\varepsilon_\pi-\omega^{(+)}_s}
2F'\int\rho^{(+)}_s(r)\chi_\pi(r)\chi_\nu(r)\frac{dr}{r^2}.
\label{eq5}
\end{equation}
Here, $n_\nu$ ($n_\pi$) are the occupation numbers for neutrons
(protons); $r^{-1}\chi_\nu$ ($r^{-1}\chi_\pi$) are the radial
bound-state wave functions for neutrons (protons);
$\nu=n_{r,\nu},j_\nu,l_\nu$ is the set single-particle quantum
numbers; $\sqrt{3}t^{(1)}_{\nu\pi}=\langle(\nu)||Y_1||(\pi)\rangle$
is the reduced matrix element; $(\nu)=j_\nu,l_\nu$;
$(\pi)=j_\pi,l_\pi$. Note, that the ratio
$-\rho^{(+)s}_{\nu\pi}/(n_\nu-n_\pi)=c^{(+)s}_{\nu\pi}$ is the
amplitude of probability to find a certain $1^-$
neutron-(proton-hole) configuration in the collective state
$|s^{(+)},1M\rangle$.

The pole representation of the effective field, strength function,
and also nucleon-escape amplitudes is valid in the energy region of
an isolated doorway-state resonance corresponding to the respective
(quasi-stationary) collective state. As applied to the IAR with the
excitation energy higher the proton separation threshold, the
Breit-Wigner-type parametrization of the Fermi strength function
$S_F(\omega^{(-)})$ and partial proton-escape amplitudes
$M_{F,\nu}(\omega^{(-)})$ are explicitly given by Eq.~(\ref{eq9}) of
Ref.~\cite{Gor1}. (See also Eq.~(6) of Ref.~\cite{Moukhai}). We show
here only the representation of the Fermi effective field:
\begin{equation}
\widetilde{V}_F(r,\omega^{(-)})\rightarrow \frac{2F'}{r^2}
\frac{S^{1/2}_A\rho_A(r)}{\omega^{(-)}-\omega^{(-)}_A+
i\ds\frac{\Gamma}{2}}. \label{eq6}
\end{equation}
Here $\rho_A(r)=\sqrt{4\pi}r^2\rho_A(\vec r)$ is the IAS transition
density; $S^{1/2}_A=\sqrt{4\pi}\int\rho_A(r)dr$, $S_A$ is the IAS
Fermi strength; $\Gamma=\sum_\nu \Gamma_\nu$ is the IAR total proton
width, while $\Gamma_\nu$ is the partial width for IAR proton decay
into the neutron-hole state $\nu^{-1}$ of the product nucleus. Note
that the ``ideal''-IAS transition density $\rho^{id}_A(\vec
r)=(N-Z)^{-1/2}n^{(-)}(\vec r)$ is proportional to the neutron
excess density. ($S^{id}_A=(N-Z)$, the neutron excess is related to
the parent nucleus).

\subsection{$(\gamma p)$-reaction amplitude}
The expression for the amplitude of the $(\gamma p)$-reaction,
accompanied by DIAR excitation, is derived in three steps. First, we
derive the CRPA expression for the amplitude of the IAR partial
proton width, $(\Gamma_\nu)^{1/2}$. This expression can be obtained
from Eqs.~(8), (9) of Ref.~\cite{Gor1} (or from Eqs.~(5), (6) of
Ref.~\cite{Moukhai}) and the pole representation of the Fermi
effective field of Eq.~(\ref{eq6}):
\begin{equation}
(\Gamma_\nu)^{1/2}=\sqrt{2\pi}N_\nu^{1/2}\delta_{(\pi)(\nu)}
\int\chi_{\varepsilon(\pi)}(r)v^{tr}_A(\vec r)\chi_\nu(r)dr.
\label{eq7}
\end{equation}
Here, $N_\nu=(2j_\nu+1)n_\nu$ is the number of neutrons filling
level $\nu$ in a parent-nucleus state;
$r^{-1}\chi_{\varepsilon(\pi)}(r)$ is the normalized to the
$\delta$-function of energy radial continuum-state (real) wave
function for the escaping proton with the energy
$\varepsilon=\varepsilon_\nu+\omega^{(-)}_A$; $v^{tr}_A(\vec r)=
2F'\rho_A(\vec r)$ is the IAS transition potential. For the
``ideal'' IAS $\omega^{(-)}_A=\Delta_C$, and the transition
potential $(v^{tr}_A)^{id}=(N-Z)^{1/2}v$  is proportional to the
mean-field isovector part, provided that the isospin-selconsistency
condition is fulfilled: $v(r)=2F'n^{(-)}(\vec r)$, where $v(r)$ is
the symmetry potential (see, e.g., Ref.~\cite{Gor1}).

The above-outlined CRPA expression of Eq.~(\ref{eq7}) can be applied
to the IAS based on the ``valence neutron + closed-shell core''
parent-nucleus state. (In such a case $N_\nu=1$). This point allows
us to generalize Eq.~(\ref{eq7}) for the case of the collective
neutron-(proton-hole)-type parent states $|s^{(+)},1M\rangle$.
Considering the IAS as the ``ideal'' one, we can substitute the
product $N_\nu^{1/2}\chi_\nu(r)$ in Eq.~(\ref{eq7}) by the sum
$\sum_\nu c^{(+)s}_{\nu\pi}\chi_\nu(r)n_\pi^{(1/2)}$ to get the
expression for the doubly-partial amplitude of the DIAR proton
width:
\begin{equation}
(\Gamma^{s}_{(\pi')\pi})^{1/2}=\sqrt{2\pi} (2T_>)^{-1/2}
n_\pi^{1/2}t^{(1)}_{(\pi')(\pi)}\int\chi_{\varepsilon(\pi')}(r)
v(r)g_{(\pi')}(r,r',\varepsilon_\pi+\omega^{(+)}_s)
2F'\rho^{(+)}_s(r')\chi_\pi(r')\frac{drdr'}{r'^2}. \label{eq8}
\end{equation}
Here, $\varepsilon=\varepsilon_\pi+\omega^{(+)}_s+\Delta_C$ is the
energy of the escaping proton, the symmetry potential $v(r)$ is
defined for the $(Z-1,N+1)$ parent nucleus,
$(rr')^{-1}g_{(\pi')}(r,r',\varepsilon')$ is the radial neutron
Green function with $(\nu)=(\pi')$. In derivation of Eq.~(\ref{eq8})
we use the relation between the probability amplitude
$c^{(+)s}_{\nu\pi}$ and corresponding transition-density element of
Eq.~(\ref{eq5}) and, also, the spectral expansion for the radial
neutron Green function. The partial width for DIAR proton decay into
the proton-hole state $(\pi)^{-1}$ of the product nucleus $(Z-1,N)$
equals $\Gamma^s_\pi=\sum_{(\pi')}\Gamma^s_{(\pi')\pi}$. Here, the
sum is taken over the proton continuum-state quantum numbers
$(\pi')=j_{\pi'},l_{\pi'}$, which are compatible with the selection
rules for the dipole transitions. It should be noted, that the use
of  Eq.~(\ref{eq7}) for derivation of Eq.~(\ref{eq8}) means, in
fact, the use a perturbation theory on  DIAR coupling to the
single-proton continuum.

Using the expression of Eq.~(\ref{eq8}) for the doubly-partial DIAR
proton-decay amplitude, one can obtain the CRPA-based expression for
the corresponding doubly-partial $(\gamma p)$-reaction amplitude
$R_{(\pi')\pi}(\omega)$. For this purpose, we use the pole
representation of Eq.~(\ref{eq4}) for the dipole effective field and
dipole strength function and also the relation of Eq.~(\ref{eq2})
between the GDR$_>$ and GDR$^+$ strength functions. The result is:
\begin{equation}
R_{(\pi')\pi}(\omega)=(B\omega)^{1/2}(2T_>)^{-1}n^{1/2}_\pi
t^{(1)}_{(\pi')(\pi)}\int\chi_{\varepsilon(\pi')}(r)v(r)
g_{(\pi')}(r,r',\varepsilon_\pi+\omega^{(+)}_C)
\widetilde{V}(r',\omega^{(+)}_C)\chi_\pi(r')drdr', \label{eq9}
\end{equation}
where  $\varepsilon=\varepsilon_\pi+\omega$. The pole representation
of this amplitude is consistent with Eq.~(\ref{eq8}):
\begin{equation}
R_{(\pi')\pi}(\omega)\rightarrow \sqrt{\frac{3\pi^2}{2}} \frac{\hbar
c}{\omega_s}\frac{(\Gamma_{\gamma_0,s})^{1/2}
(\Gamma^{s}_{(\pi')(\pi)})^{1/2}}{\omega-\omega_s+i0}. \label{eq10}
\end{equation}
Being based on Eqs.~(\ref{eq9}), (\ref{eq10}), and the expression
for the DIAR elastic radiative width (Subsect.~\ref{diar_prop}), we
can evaluate within the CRPA the partial proton widths for a given
DIAR. An expression similar to Eq.~(\ref{eq9}) has been derived in
Ref.~\cite{Rumyan} in an intuitive way.

\section{\label{diar}Properties of the DIAR in $^{90}$Zr}
\subsection{Input quantities and model parameters}
A nuclear mean field and particle-hole interaction are the input
quantities for any CRPA-based approach. Along with the
phenomenological (non-spin-flip and momentum-independent)
Landau-Migdal particle-hole interaction, we use a realistic
phenomenological isoscalar part  of the nuclear mean field
(including the spin-orbit term), while the mean-field isovector part
(the symmetry potential) and also the mean Coulomb field are
calculated selfconsistently (see, e.g. Refs.~\cite{Gor1,Gor2}). The
parametrization of the mean-field isoscalar part is explicitly given
in Ref.~\cite{Gor2}. To describe better the neutron
single-quasi-particle spectrum in $^{89,91}$Zr, the strength
parameters are somewhat specified:
\begin{equation}
U_0=54.9\mbox{MeV},\qquad U_{S0}=13.85\mbox{MeV\,fm$^2$},\qquad
f'=1.03, \label{eq11}
\end{equation}
where $F'=f'\,300$ MeV fm$^3$.

Following Ref.~\cite{Safonov}, we take effectively into account the
contribution of the isovector momentum-dependent forces in formation
of the GDR$^{(\pm)}$ by the scaling transformation of the CRPA
strength functions: $S^{(\pm)}(\omega^{(\pm)})\rightarrow
(1+\widetilde{k}')^{-1}S^{(\pm)}(\omega^{(\pm)})/(1+\widetilde{k}'))$.
The parameter $\widetilde{k}'$ describes the relative contribution
of the momentum-dependent forces to the corresponding
energy-weighted sum rule. For $^{90}$Zr the $\widetilde{k}'=0.1$
value has been found in Ref.~\cite{Safonov} from independent data
\cite{Gor3}. In the properly modified CRPA calculations of the
photo-absorption cross section $\sigma^>_a(\omega)$ (see also
Subsect.~\ref{diar_prop}), performed in Ref.~\cite{Safonov} for
$^{48}$Ca and $^{90}$Zr, a few isolated DIAR have been found.

To improve description of the DIAR in $^{90}$Zr, in the present work
we take proton pairing approximately into account by simply
substituting in CRPA equations the occupation numbers $n_\pi$ by the
Bogolyubov coefficients $v^2_\pi$. The same modification is used to
calculate the neutron-excess density in order to keep the
isospin-selfconsistency condition ~\cite{Gor1,Rodin}. The energy gap
parameter is taken 1.0 MeV  (this value is close to that used in
Ref.~\cite{Rodin}), while the proton chemical potential is properly
calculated to find, finally, the coefficients $v^2_\pi$.

\subsection{Calculation results}
\begin{figure}
\includegraphics[scale=1.0]{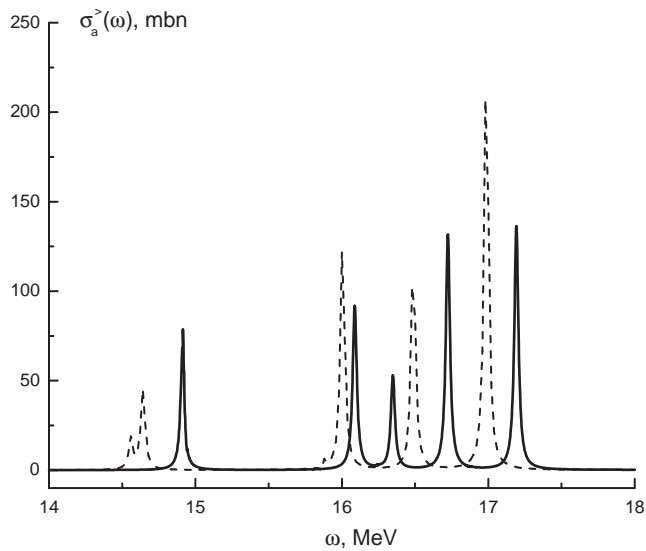}
\caption{The cross sections of photo-absorption, accompanied by
excitation of the isolated DIAR in $^{90}$Zr. The full and dotted
lines correspond, respectively, to calculations with and without
taking the proton pairing into account.\label{Fig1}}
\end{figure}

We calculate first the photo-absorption cross section
$\sigma^>_a(\omega)$ for the excitation energy interval 14--18 MeV,
which corresponds to the low-energy part of the GDR$^{(+)}$ strength
function of Eq.~(\ref{eq2})~\cite{Safonov}. In strength function
calculations (performed with and without taking proton pairing into
account) we add an artificial small imaginary part to the
single-particle potential, used within the CRPA for calculations of
the $\omega$-dependent quantities, to make presentation of the
results more convenient (Fig.~\ref{Fig1}). Within the mentioned
interval we find a few DIAR, whose energies $\omega_s$  and
calculated via $(\sigma^>_a){}^{int}_s$ radiative widths
$\Gamma_{\gamma_0,s}$ are given in Table~\ref{tab1}. The
particle-hole part of the DIAR transition density
$\rho_{>,s}(r)=(2T_>)^{(-1/2)}\rho^{(+)}_s(r)$ is next calculated
via the residues at the poles in the energy dependence of the dipole
effective field of Eq.~(\ref{eq4}). The calculated transition
densities are shown in Fig.~\ref{Fig2}. These densities are used to
evaluate within the improved plane wave approximation the DIAR form
factor of Eq.~(\ref{eq3}). The results are shown in Fig.~\ref{Fig3}.
Further, Eqs.~(\ref{eq9}), (\ref{eq10}) are used to evaluate the
DIAR partial proton widths $\Gamma_{\pi_0-\pi_3}$ for population of
the $2p_{1/2}$ ($\pi_0$), $1g_{9/2}$ ($\pi_1$), $2p_{3/2}$
($\pi_2$), $1f_{5/2}$ ($\pi_3$) one-hole states in $^{89}$Y. The
calculated widths are given in Table~\ref{tab1} together with the
experimental data of Ref.~\cite{Cosel}, related to the certain DIAR
($E_x=16.1$ MeV). For this DIAR we show also the calculated
doubly-partial proton widths $\Gamma^s_{(\pi')\pi}$ for decay into
the main channels $(\pi_0)$ and $(\pi_2)$ (Table~\ref{tab2}).

Using the calculated DIAR total proton width
$\Gamma_s=\sum_\pi\Gamma^s_\pi$ we can evaluate the $(\gamma
p_{tot})$-reaction cross section for the considered
excitation-energy interval:
\begin{equation}
\sigma^>_{\gamma p_{tot}}(\omega)=\frac{1}{2\pi}
\sum_s\frac{(\sigma^>_a){}^{int}_s\Gamma_s}
{(\omega-\omega_s)^2+\ds\frac{1}{4}\Gamma_s^2}. \label{eq12}
\end{equation}

To compare with the experimental data of Ref.~\cite{Rodin}, we make
``apparatus'' averaging this cross section with the function
$I(\omega,\omega')=(I/\pi)/((\omega-\omega')^2+I^2)$. The averaged
cross sections, obtained for the appropriate $I$ values ($I=50$ and
100 keV), are shown in Fig.~\ref{Fig4}. together with the
experimental data of Ref.~\cite{Shoda}. In calculations of the
mentioned cross section the relatively small contribution of the
GDR$_<$ excitation is also taken into account. It is done within the
similar approach by the methods described in Ref.~\cite{Gor3}].

\subsection{Discussion of the results} 
\begin{figure}
\includegraphics[scale=1.0]{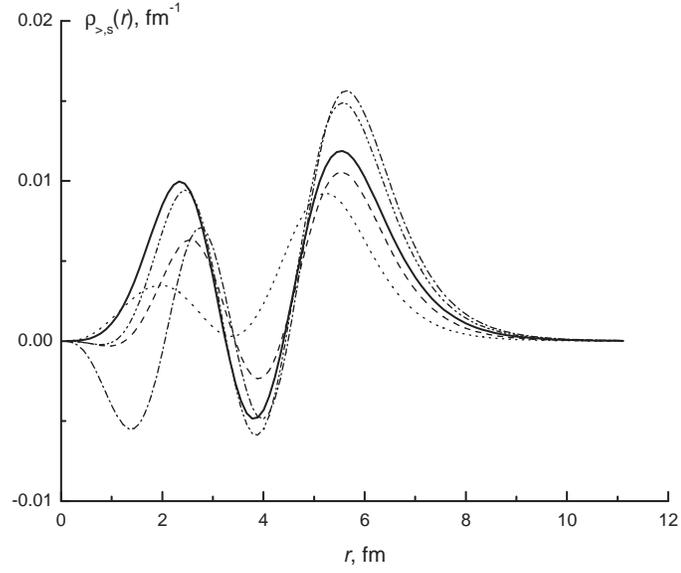}
\caption{The calculated proton-(proton-hole) part of the DIAR
transition density. The dashed, full, pointed, dashed-dotted and
dashed-dot-dotted lines correspond to the DIAR at 14.9, 16.1, 16.3,
16.7, 17.2 MeV, respectively.\label{Fig2}}
\end{figure}

\begin{figure}
\includegraphics[scale=1.0]{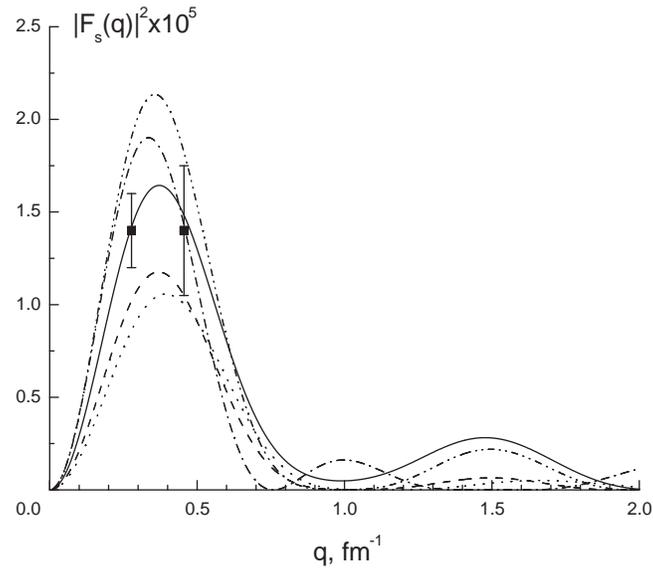}
\caption{The DIAR form factors, calculated within the improved plane
wave approximation. The notations are the same, as in
Fig.~\ref{Fig2}. The experimental data concerned with the DIAR at
16.1 MeV are taken from Ref.~\cite{Cosel}.\label{Fig3}}
\end{figure}

\begin{figure}
\includegraphics[scale=1.0]{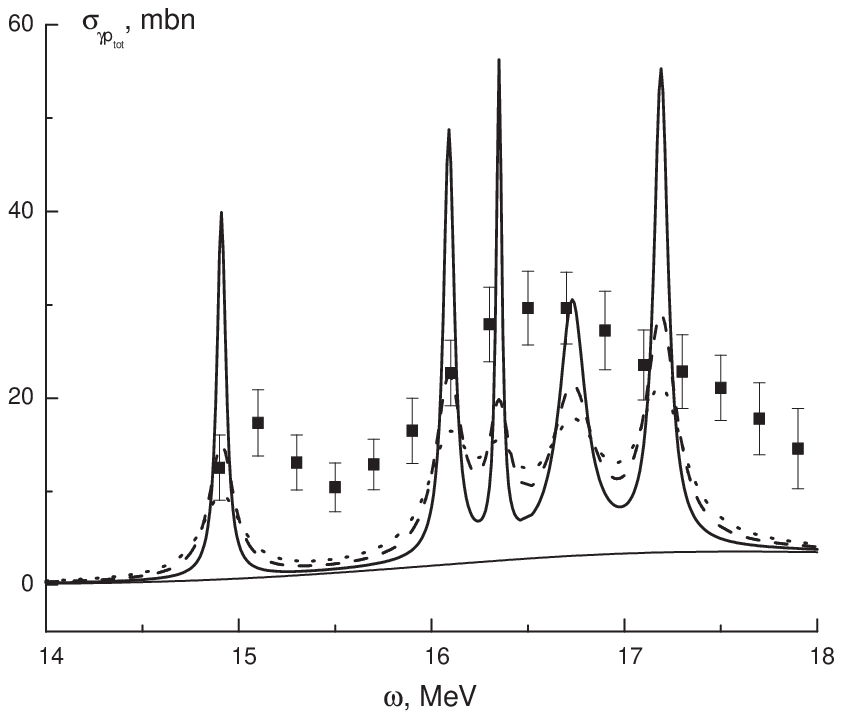}
\caption{The $^{90}$Zr$(\gamma p_{tot})$-reaction cross section,
calculated for the excitation-energy interval 14--18 MeV (full
line). The dashed and dotted correspond to averaging with the
parameter $I$ equal 50 and 100 keV, respectively. The thin line is
for GDR$_<$ contribution to the cross section. The experimental data
are taken from Ref.~\cite{Shoda}.\label{Fig4}}
\end{figure}

Due to the
specific shell-model structure of $^{90}$Zr, the proton pairing in
this nucleus is rather weak. In particular, the number of protons,
occupying the $\pi$ level, $N_\pi=(2j_\pi+1)$ $v_\pi^2$, is
relatively small for the $1g_{9/2}$ level. (In Table~\ref{tab2} the
$N_\pi$ values are shown for a few levels near the proton chemical
potential). Nevertheless, due to pairing, the DIAR with the energy
16.3 MeV appears in calculations (Fig.~\ref{Fig1}) and DIAR proton
decay into the $9/2^+$-state of $^{89}$Y becomes formally possible
(Table~\ref{tab1}). Among the IAR found in calculations, three of
them have relatively large elastic radiative width
(Table~\ref{tab1}), transition density (Fig.~\ref{Fig2}) and,
therefore, the form factor (Fig.~\ref{Fig3}). The DIAR with the
energy 16.1 MeV can be related to the resonance experimentally
studied in Ref.~\cite{Cosel}. Within the present approach we
obtained a satisfactory description of the decay properties of this
DIAR (Table~\ref{tab1}). The $d_{3/2}$ component of the proton
partial widths is found to be the main one for decay into the
$\pi_0$-decay channel. This observation is in agreement with the
analysis of the DIAR wave function performed in Ref.~\cite{Cosel}.
However, it is not truth for another main decay channel $(\pi_2)$ as
it follows from Tables~\ref{tab1} and~\ref{tab2}. The two-node
transition densities of all DIAR (Fig.~\ref{Fig2}) show that the
low-energy $1^-$ parent states are not strongly collectivized, as it
expected for the pygmy resonance. (The main-peak GDR$^{(+)}$
transition density is nodeless). The form factor, calculated for the
DIAR with the energy 16.1 MeV, is in reasonable agreement with
experimental data of Ref.~\cite{Cosel}. The calculation results show
that at least  two DIAR (with the calculated energy 16.7 and 17.2
MeV) can be found experimentally via the $(e,e'p)$-reaction. This
statement is partially supported by comparing the calculated and
experimental $(\gamma p_{tot})$-reaction cross section within the
excitation-energy interval 14--18 MeV (Fig.~\ref{Fig4}). It is
noteworthy that the approximations used in description of the IAS,
proton pairing, and single-particle level scheme lead to
uncertainties in evaluation of the DIAR energies (presumably, in a
few hundreds keV). Description of the DIAR decay properties seems to
be not much sensitive to these uncertainties.

In conclusion, we first present continuum-RPA-based description of
the decay properties of the specific doubly-collective states,
$1^{-}$-IAR, and describe satisfactorily the available experimental
data.

\section*{Acknowledgement}

The authors would like to thank M.N.~Harakeh and
P.~von~Neumann-Cosel for interesting discussions and valuable
remarks. One of the authors (M.H.U.) thanks M.N.~Harakeh for warm
hospitality during a long-term stay at KVI and acknowledges the
support from the ``Nederlandse organisatie voor wetenschappelijk
onderzoek'' (NWO). This work is partially supported by the Russian
Fund for Basic Researches (RFBR) under the grant
No.~06-02-016883-a.

\begin{table}
\caption{The energies (in MeV), elastic radiative and partial
protons widths (in eV and keV, respectively), calculated for the
isolated DIAR in $^{90}$Zr. The parameters deduced in
Ref.~\cite{Cosel} for the observed DIAR  (the widths are given with
errors) are also shown.\label{tab1}}
\begin{tabular}{cccccc}
\hline\hline
 $\omega_s$ & $\Gamma_{\gamma_0}$ & $\Gamma_{\pi_0}$ & $\Gamma_{\pi_1}$ & $\Gamma_{\pi_2}$ & $\Gamma_{\pi_3}$ \\
\hline
 14.91 & 68  & 49   & 0.1 & 8    & --   \\
 16.09 & 110 & 55.1 & 1.7 & 10.6 & 0.12 \\
 16.28 & $108\pm35$ & $50\pm6$ & $< 2.0$ & $20\pm5$ & $7\pm2$ \\
 16.35 & 66 & 8 & 6.5 & 20 & 0.10 \\
 16.73 & 169 & 6 &0.4 & 159 & 0.02 \\
 17.19 & 185 & 8 & 0.3 & 81 & 0.06 \\
\hline\hline
\end{tabular}
\end{table}

\begin{table}
\caption{The doubly-partial widths (in keV), calculated for the main
channels of proton decay of the DIAR at 16.1 MeV in
$^{90}$Zr.\label{tab2}}
\begin{tabular}{l|c|c|c|c|c|c|c|c|c|c|c}
\hline\hline
 $\pi$        & \mc{2}{$2p_{1/2}$} & \mc{3}{$1g_{9/2}$} & \mc{3}{$2p_{3/2}$} & \mc{3}{$1f_{5/2}$} \\
\hline
 $N_\pi$      & \mc{2}{1.49} & \mc{3}{0.83} & \mc{3}{3.77} & \mc{3}{5.89} \\
\hline
 $(\pi')$     & $s_{1/2}$ & $d_{3/2}$ & $f_{7/2}$ & $h_{9/2}$ & $h_{11/2}$ &
 $s_{1/2}$ & $d_{3/2}$ & $d_{5/2}$ & $d_{3/2}$ & $d_{5/2}$ & $h_{7/2}$ \\
\hline
 $\Gamma_{(\pi'),\pi}$ & 0.0 & 55.1 & 0.1 & 0.0 & 1.6 & 2.2 & 3.8 &
 4.6 & 0.01 & 0.10 & 0.0 \\
 \hline\hline
\end{tabular}
\end{table}


\begin{thebibliography}{12}
\bibitem{Harakeh} M.N.~Harakeh and A.~van~der~Woude, ``Giant Resonances:
Fundamental High-Frequency Modes of Nuclear Excitations'' (Oxford
University Press, New York, 2001).

\bibitem{Auerbach} N.~Auerbach and A.~Klein, Nucl. Phys. A395 (1983) 77.

\bibitem{Safonov} I.V.~Safonov, M.L.~Gorelik, and M.H.~Urin, Phys. At.
Nucl. 69 (2006) 403.

\bibitem{Cosel} P.~von~Neumann-Cosel et al., Phys. Rev. Lett. 78 (1997) 2924.

\bibitem{Moukhai} E.A.~Moukhai, V.A.~Rodin, and M.H.~Urin, Phys. Lett. B 447
(1999) 8.

\bibitem{Gor1} M.L.~Gorelik and M.H.~Urin, Phys. Rev. C63 (2001) 064312.

\bibitem{Landau} L.D.~Landau and E.M.~Lifshitz, ``Quantum Mechanics
(Non-Relativistic Theory)'', 3rd ed. (Pergamon Press, Oxford,
England, 1977).

\bibitem{Rumyan} O.A.~Rumyantsev and M.H.~Urin, Izv. Akad. Nauk. SSSR, Ser. Fiz.
55 (1991) 866.

\bibitem{Gor2} M.L.~Gorelik, S.~Shlomo, and M.H.~Urin, Phys. Rev. C62,
(2001) 044301.

\bibitem{Gor3} M.L.~Gorelik and M.H.~Urin, Phys. At. Nucl. 69 (2006) 219.

\bibitem{Rodin} V.A.~Rodin and M.H.~Urin, Phys. At. Nucl., 66 (2003) 2128.

\bibitem{Shoda} K.~Shoda et al., Nucl. Phys. A239 (1975) 397.
\end{thebibliography}
\end{document}